\documentclass[12pt]{emulateapj}

\usepackage{natbib}
\usepackage{rotating}

\shorttitle{CEP0227: pulsation modelling}
\shortauthors{Marconi, et al. }

\begin{document}


\title{The Eclipsing Binary Cepheid OGLE-LMC-CEP-0227 in the Large Magellanic
  Cloud: pulsation modelling of light and radial velocity curves}

\author{M. Marconi\altaffilmark{1}, R. Molinaro\altaffilmark{1},
G. Bono\altaffilmark{2,3}, G. Pietrzy{\'n}ski\altaffilmark{4,5}, 
W. Gieren\altaffilmark{5}, B. Pilecki\altaffilmark{4,5}, 
R. F. Stellingwerf\altaffilmark{6}, D. Graczyk\altaffilmark{5}, 
R. Smolec\altaffilmark{7}, P. Konorski\altaffilmark{4}, 
K. Suchomska\altaffilmark{4}, M. G{\'o}rski\altaffilmark{4}, 
P. Karczmarek\altaffilmark{4}}

\altaffiltext{1}{INAF-Osservatorio astronomico di Capodimonte, Via
Moiariello 16, 80131 Napoli, Italy; marcella.marconi@oacn.inaf.it, molinaro@oacn.inaf.it}
\altaffiltext{2}{Dipartimento di Fisica - Universit\`a di Roma Tor
Vergata, Via della Ricerca Scientifica 1, 00133 Roma, Italy; giuseppe.bono@roma2.infn.it}
\altaffiltext{3}{INAF-Osservatorio Astronomico di Roma, Via Frascati 33, 00040 Monte Porzio Catone, Italy} 
\altaffiltext{4}{Warsaw University Observatory, Al. Ujazdowskie 4, 00-478, Warsaw,
Poland; pietrzyn@astrouw.edu.pl, pilecki@astrouw.edu.pl,
piokon@astrouw.edu.pl, ksenia@astrouw.edu.pl, mgorski@astrouw.edu.pl, pkarczmarek@astrouw.edu.pl}
\altaffiltext{5}{Universidad de Concepci{\'o}n, Departamento de Astronomia, Casilla 160-C,
Concepci{\'o}n, Chile; wgieren@astro-udec.cl, darek@astro-udec.cl}
\altaffiltext{6}{Stellingwerf Consulting, 11033 Mathis Mtn Rd SE, 35803 Huntsville, AL USA; rfs@swcp.com}
\altaffiltext{7}{Nicolaus Copernicus Astronomical Centre, Bartycka 18, 00-716 
Warszawa, Poland; smolec@camk.edu.pl}

\begin{abstract}
We performed a new and accurate fit of light and radial velocity curves 
of the Large Magellanic Cloud (LMC) Cepheid --OGLE-LMC-CEP-0227-- belonging 
to a detached 
double-lined eclipsing binary system. We computed several sets of nonlinear, 
convective models covering a broad range in stellar mass, effective temperature 
and in chemical composition. The comparison between theory and observations 
indicates that current theoretical framework accounts for luminosity 
--V and I band-- and radial velocity variations over the entire pulsation cycle. 
Predicted pulsation mass --$M=4.14\pm0.06M_{\odot}$-- and mean effective 
temperature --$T_e$=6100$\pm$50 K-- do agree with observed estimates with
an accuracy better than 1$\sigma$.  The same outcome applies, on
average, to the luminosity amplitudes and to the mean radius. 
We find that the best fit solution requires a chemical composition that 
is more metal--poor than typical LMC Cepheids (Z=0.004 vs 0.008) and 
slightly helium enhanced (Y=0.27 vs 0.25), but the sensitivity to He 
abundance is quite limited.  
Finally, the best fit model reddening --$E(V-I)=0.171\pm0.015$ mag-- 
and the true distance modulus corrected  for the       
barycenter of the LMC  --$\mu_{0,LMC}$=18.50$\pm$0.02$\pm$0.10
(syst) mag--, agree quite well with similar estimates in the
recent literature. 
\end{abstract}  



\section{Introduction}
Classical Cepheids are among the most used primary distance
indicators, currently at the base of an accurate extragalactic
distance scale \citep[]{f01,s01, f11}, thanks to their
characteristic Period-Luminosity (PL) relation.
At the same time, these intermediate mass (typically from 3 to 13
$M_{\odot}$) helium burning pulsating stars are used as tracers of
relatively young ($\sim$ 10 $\div$ 100 Myr) stellar populations.
The physical basis of the PL relation is well known.  By combining 
the period-density and the Stephan-Boltzmann relations, 
 we obtain a Period-Luminosity-Color-Mass (PLCM) 
relation. Stellar evolution predicts the occurrence of a 
Mass Luminosity (ML) relation for intermediate-mass He burning stars 
\citep[][]{b00} and the PLCM can be easily 
converted into a Period-Luminosity-Color (PLC) relation holding for 
each individual Cepheid. The projection of the PLC onto the PL plane 
gives the PL relation.
On this basis, the investigation of Cepheid pulsation properties is
also crucial for providing independent constraints on the ML
relation. 

In this context one of the most popular open problem is the 
so-called mass discrepancy problem suggested by 
\citet{Christy70} and \citet{Stobie69}. 
The quoted authors noted  that the Cepheid masses evaluated from the
comparison of theoretical isochrones with
observations were larger by almost a factor
of two  than the masses obtained from application of a 
period-mass-radius relation \citep{Fricke71}. 
Even if a  reduction of this discrepancy between evolutionary and pulsational
masses was obtained by \citet{Moskalik92} on the basis of updated
radiative opacities, namely by the  OPAL
\citep{ir91} and the Opacity Project \citep{sea94} groups, several
authors \citep{c05,n08,bono02,kw06, Evans2005,bro03}
noted tha
the effect remained of the order of 10$\%$-20$\%$ for both Galactic
and Magellanic Cepheids.

A crucial input to this scenario was provided by the recent detection
of two Cepheids in well detached, double-lined eclipsing binary
systems in the LMC \citep[][OGLE-LMC-CEP-0227, OGLE-LMC-CEP-1812]{pie10,pie11}. 
The geometry and the precision of both photometric and 
spectroscopic data allowed the authors to measure the masses
of these Cepheids with the unprecedented precision of 1$\%$, thus
fixing a cornerstone in the solution of the mass discrepancy
problem. Indeed, \citet[][]{cs11} were able to find evolutionary
masses consistent with the dynamical estimate for OGLE-LMC-CEP-0227,
on the basis of stellar evolution models that include a
moderate amount of convective core overshooting,
 On the other hand, \citet{n11} found that both moderate convective core overshooting and pulsation-driven 
mass loss are required  to achieve a solution of the mass
 discrepancy problem and\citet[][]{pra12} adopt a new Bayesian approach to 
constrain the intrinsic stellar parameters of OGLE-LMC-CEP-0227, in
particular its mass and age, by varying the efficiency 
of mass loss and core overshooting.

In this investigation we constrain the pulsation properties of 
OGLE-LMC-CEP-0227, by modelling its light and radial velocity curves with 
nonlinear convective models specifically computed for this purpose. 
In \S2 we present the observational and theoretical framework, 
while in \S3 we discuss the procedure adopted to fit light and 
velocity curves. In the same section, we also investigate the 
impact of metal and helium content on  the best fit models. 
Conclusions and final remarks close the paper.

\section{Observational and theoretical framework}

The observables we plan to use in the comparison between theory and observations 
are the V and I-band light curves (with the amplitudes $A_V$, $A_I$)  and the 
radial velocity curve. Note that we define the light and the radial velocity 
curves {\em observables} even if they have been estimated by deblending the 
cumulative light and radial velocity curves of the binary system.  
To further constrain the projection factor, i.e. the parameter adopted to 
transform radial velocity into pulsation velocity, the radial velocity 
amplitude in our approach was assumed as a free parameter. The above 
observables together with the pulsation periods are available for a 
large number of classical Cepheids, in particular for those observed 
with one of the different flavors of the Baade--Wesselink method.  

During the last few years it has been found that nonlinear, convective 
Cepheid models can play a crucial role in fitting the quoted observables 
\citep{bms99,bms00,bono02,kw06}.
These models, once fixed the ML relation, provide very 
accurate fits of the above observables. On the basis of this 
theoretical framework and by assuming plausible values for the 
chemical composition (metal --Z-- and helium --Y-- abundances) 
and for the effective temperature very accurate fits have been 
provided not only for Galactic and Magellanic Cloud (MC) Cepheids 
\citep[see e.g.][]{was97,bono02,kw06,n08,m13}, but also 
for other groups of radial variables \citep{marconi09}. 
A further advantage of this nonlinear pulsation approach is that 
it provides independent estimates of the individual distance modulus 
and of the reddening \citep{bono02,mc05,md07,m13}.

Moreover, intrinsic parameters of radial variables based on pulsation 
observables can be compared with those based on evolutionary observables.   
However, this approach is partially hampered by the fact that pulsation 
models are envelope models, and therefore, they require the mass-luminosity 
relation predicted by evolutionary models. The binary LMC Cepheid OGLE-LMC-CEP-0227 
provides, for the first time, the opportunity to break this degeneracy.  
The unprecedented accuracy of the dynamical mass allows us to use a restricted 
range in mass values. Moreover, we also have for the same object a precise 
estimate of the mean effective temperature, this means that we can change 
the mean luminosity until predicted and observed period agree with each 
other. We still lack detailed information concerning the chemical composition, 
but for LMC Cepheids spectroscopic investigations are already available in the 
literature (Romaniello et al. 2008).  
The observed parameters of OGLE-LMC-CEP-0227 are listed in the first row of Table~1
\footnote{The mean radius listed in column 11 is slightly different 
than the value given in Pietrzynski et al. (2010). The new estimate plus the 
projection factor --p-- are based on the results by Pilecki et al. (2013, 
in preparation).}.

\section{Fit of light and velocity curves}

To perform an accurate fit of the pulsation observables --V and I-band
light and radial velocity curves, we constructed a large set
of pulsation models by adopting the typical chemical composition of
LMC (Z=0.008, Y=0.25) young stellar population
\citep[][]{l98,r08,mu11}, the dynamical mass measured by \citet{pie10}
and a broad range of effective temperatures with a step of 50 K. The
luminosity of individual models for each value of the effective
temperature was changed in order to match the pulsation
period\footnote{The exact match between observed and predicted 
period was fixed by using linear nonadiabatic pulsation models that 
provide the static envelope structure to the nonlinear hydrodynamical models 
\citep[][]{bms99}. The period predicted by nonlinear models at limit cycle 
stability can differ from the linear one by a few percent.} (see Table~1). 
Fig.~1 shows the comparison between a subset of the computed models
and the observations for OGLE-LMC-CEP-0227. Data plotted in this figure  
indicate that theory agrees quite well with observations.   

\begin{figure*}
\includegraphics[height=0.8\textheight,width=0.8\textwidth]{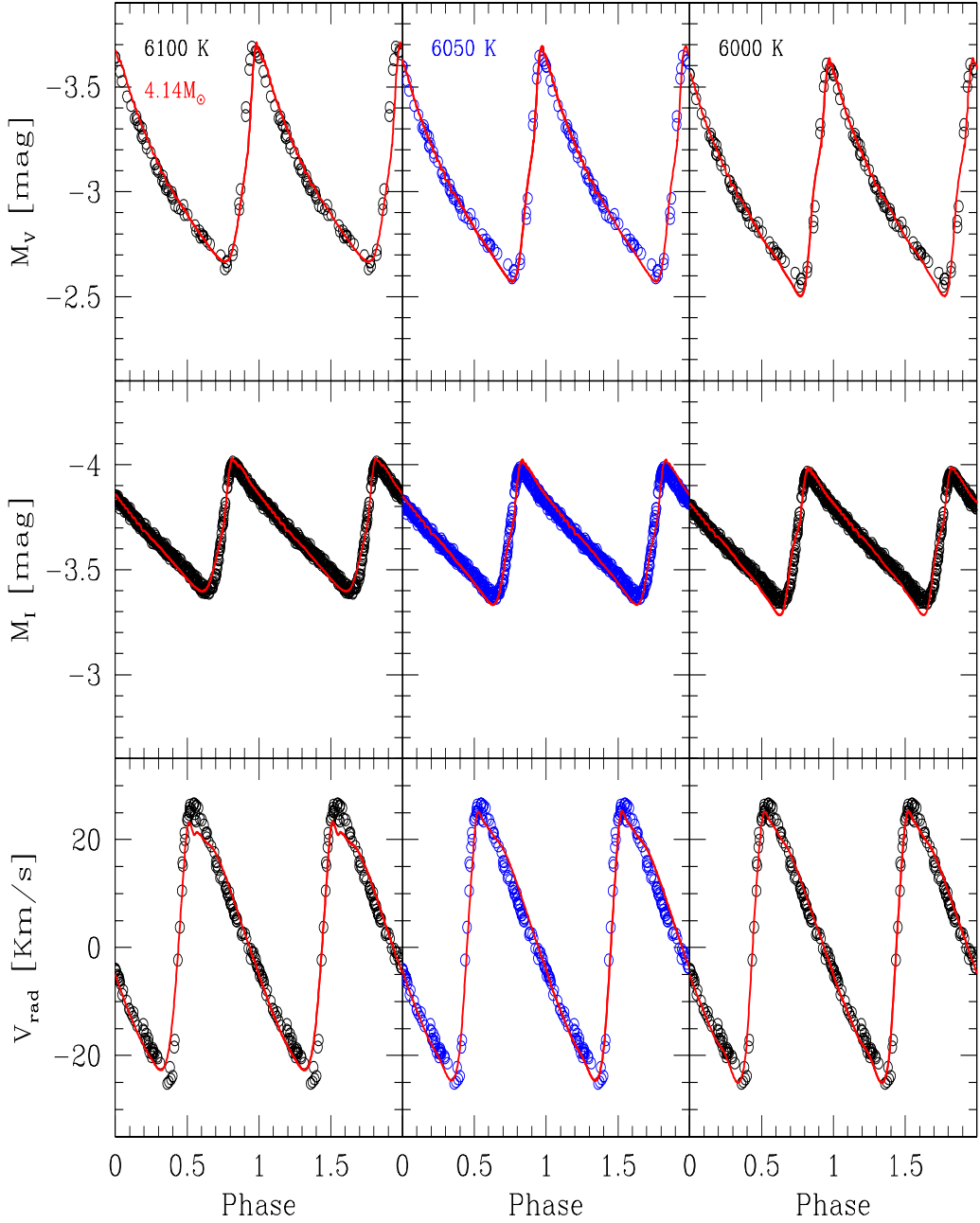} \label{fig1}  
\caption{From top to bottom V- and I-band light curves and radial
  velocity curve. Open circles mark observations, while red lines
  display nonlinear pulsation models constructed at fixed stellar mass
  (M=4.14$M_{\odot}$) and chemical composition  (Z=0.008,
  Y=0.25). From left to right the different panels show pulsation
  models constructed assuming different values of the mean effective
  temperature (labeled values). The luminosity of individual models
  was changed until predicted and observed period agreed with each
  other. The blue open circles mark the best fit solution, according
  to the total  rms, for this set of models. The apparent magnitudes
  of the light curves were transformed into absolute magnitude using
  the true distance modulus and the reddening listed in Table~1.  
}
\end{figure*}

However, to constrain on a quantitative basis the quality of the fit between 
the different sets of models and observations, we adopted a $\chi^2$ analysis. 
We phased the light and pulsational velocity curves of models in order 
to locate the maximum of the V-band light curve at phase
zero. Moreover, the best  fit between theory and observations was
computed by minimizing the following  
$\chi^2$ functions:
\small{
\begin{eqnarray}
\chi_{Ph}^2=\sum_{i=0}^{N_{band}}\sum_{j=1}^{N_{points}}\left [m^i_j -\left 
(M^i_{mod}\left (\phi^i_j+\delta \phi^i\right )+\delta M^i\right
  )\right ]^2\\
\chi_{VR}^2=\sum_{j=1}^{N_{points}}\left [v^{rad}_j -
\left (-\frac{1}{p}V_{mod}\left (\phi_j+\delta \phi_{VR}\right
)+\gamma \right )\right]^2 
\end{eqnarray}
}
From top to bottom they refer to light  and radial velocity curves. 
Note that the index $i$ runs over the two photometric bands and $j$ over 
the number of measurements, $N_{points}$= 68, 420, 135 for
V, I and radial velocity, respectively. The predicted 
light and pulsational velocity curves are indicated with  
$M_{mod}$ and $V_{mod}$ and they have been evaluated 
at the same phase --$\phi_j$-- of the j$^{th}$ measurement 
--$m_j$, $v^{rad}_j$-- by fitting with a spline the predicted 
curves. The magnitude $\chi^2$ function is minimized 
with respect to shifts in $\delta M$  and to 
the possible shifts in phase --$\delta \phi$-- 
(typically ranging from 0.01 to 0.02 for the models listed 
in Table~1), while the velocity $\chi^2$ function is minimized with 
respect to the $\delta \phi$ (the same as for the photometry at the
0.01 level), the barycentric velocity --$\gamma$-- 
and the projection factor $p$, i.e. the parameter adopted to convert 
the observed radial velocity into pulsational velocity. 
The phase difference applied to light and radiative velocity curves 
are quite similar, and indeed we found that both the observed and the
predicted  phase lag, defined by using Fourier phases, between V-band
light and radial velocity curves is -0.2.   
It is worth noting that the shifts in $\delta M$ provide the Cepheid true 
distance modulus and the reddening, while $\gamma$ gives 
the barycentric velocity of the system.  
 
We neglected the models for which the $\chi^2$ analysis provided 
a shift in radial velocity --$\gamma$-- which are not consistent,
within the errors, with the observed mean barycentric velocity 
(see Table~1).  
The rms for the different observables, given by the $\chi^2$ analysis, 
have been normalized to their pulsation amplitudes and summed in 
quadrature to obtain a total rms. 

The best fit solution, for each set of models, satisfy the above 
requirements concerning the shifts and attains the smallest total 
rms. The best fit model is plotted with open blue circles in Fig.~1 
and the intrinsic parameters are listed in Tab.\ref{tab-bestFit} 
(see discussion below).

\begin{table*}
{\scriptsize 
\begin{center}
\caption{The first line give the parameters derived from the
  dynamical analysis by \citet{pie10} and those of the observed
  photometric and radial velocity curves, while the best fit model
  parameters are listed in the following seven lines of the table:
  Metallicity (column 1), Helium content (column 2), Mass in solar
  units (column 3), Luminosity (column 4), Effective Temperature
  (column 5), Period (derived from the linear and the nonlinear analysis) in days (column 6),  Amplitudes of the
  photometric curves expressed in magnitudes (columns 7, 8),
  V-I Reddening value in magnitudes (column 9), dereddened Distance
  Modulus in magnitudes (column 10), the mean Radius (column 11),
  Projection Factor (column 12), Barycentric Velocity in km/s (column
  13) and Total RMS (column 14).}    
\begin{tabular}{c @{} c @{} c @{} c @{} c @{} c @{} c @{} c @{} c @{}
    c @{} c @{} c @{} c @{} c @{}}
\hline
\hline
Z &\hspace{0.3cm}Y &\hspace{0.3cm} M &\hspace{0.3cm}
$\log\left(L/L_\odot\right)$ &\hspace{0.3cm} T$_e$ &\hspace{0.3cm} P
&\hspace{0.3cm} A$^{mod}_V$ &\hspace{0.3cm} A$^{mod}_I$
&\hspace{0.3cm} E(V-I) &\hspace{0.3cm} $\mu_0$ &\hspace{0.3cm} R
&\hspace{0.3cm} p &\hspace{0.3cm} $\gamma$ &\hspace{0.3cm} rms\\  
 &\hspace{0.3cm}  &\hspace{0.3cm} $(M_\odot)$ &\hspace{0.3cm} $(dex)$
&\hspace{0.3cm} $(K)$ &\hspace{0.3cm} $(days)$  &\hspace{0.3cm} $(mag)$
&\hspace{0.3cm} $(mag)$ &\hspace{0.3cm} $(mag)$  &\hspace{0.3cm} $(mag)$
&\hspace{0.3cm} $(R_\odot)$  &\hspace{0.3cm}  &\hspace{0.3cm} $(km/s)$
&\hspace{0.3cm} \\ 
\hline &\hspace{0.3cm}  &\hspace{0.3cm} 4.14 &\hspace{0.3cm}
&\hspace{0.3cm} 5900 &\hspace{0.3cm} 3.793 &\hspace{0.3cm} 1.06 
&\hspace{0.3cm} 0.63 &\hspace{0.3cm} &\hspace{0.3cm} &\hspace{0.3cm}
33.7 &\hspace{0.3cm} 1.16 
&\hspace{0.3cm} 256.70 &\hspace{0.3cm}\\    
&\hspace{0.3cm}  &\hspace{0.3cm} $\pm$0.05 &\hspace{0.3cm} 
&\hspace{0.3cm} $\pm$250 &\hspace{0.3cm} &\hspace{0.3cm} $\pm$0.02
&\hspace{0.3cm} $\pm$ 0.01 &\hspace{0.3cm}  &\hspace{0.3cm}
&\hspace{0.3cm}$\pm$1.5 &\hspace{0.3cm}$\pm$0.14 &\hspace{0.3cm} $\pm$0.10
&\hspace{0.3cm}\\     
\hline0.008 &\hspace{0.3cm} 0.25 &\hspace{0.3cm} 4.14 &\hspace{0.3cm}
3.13 &\hspace{0.3cm} 6050 &\hspace{0.3cm} 3.793-3.805 &\hspace{0.3cm} 
1.11 &\hspace{0.3cm} 0.69 &\hspace{0.3cm} 0.154
&\hspace{0.3cm} 18.45 &\hspace{0.3cm} 33.9 &\hspace{0.3cm} 1.08
&\hspace{0.3cm} 256.80 
&\hspace{0.3cm} 0.077 \\     
 &\hspace{0.3cm}  &\hspace{0.3cm} $\pm$ 0.06 &\hspace{0.3cm} $\pm$0.02
&\hspace{0.3cm} $\pm$50 &\hspace{0.3cm} 
&\hspace{0.3cm} $\pm$0.06 &\hspace{0.3cm} $\pm$0.04 &\hspace{0.3cm}
$\pm$0.014 &\hspace{0.3cm} $\pm$0.03 &\hspace{0.3cm}
$\pm$0.3&\hspace{0.3cm} $\pm$0.05 
&\hspace{0.3cm}$\pm$0.10 &\hspace{0.3cm}\\     
\hline 0.008 &\hspace{0.3cm} 0.25  &\hspace{0.3cm} 4.08
&\hspace{0.3cm} 3.13 &\hspace{0.3cm} 6050 &\hspace{0.3cm} 3.793-3.803
&\hspace{0.3cm} 1.12 &\hspace{0.3cm} 0.69 &\hspace{0.3cm} 0.154
&\hspace{0.3cm} 18.44 &\hspace{0.3cm} 33.7&\hspace{0.3cm} 1.08
&\hspace{0.3cm} 256.78 &\hspace{0.3cm} 0.077\\     
 &\hspace{0.3cm}  &\hspace{0.3cm} $\pm$0.06 &\hspace{0.3cm} $\pm$0.02 
 &\hspace{0.3cm} $\pm$50 &\hspace{0.3cm} &\hspace{0.3cm} $\pm$0.09
&\hspace{0.3cm} $\pm$0.05 &\hspace{0.3cm} $\pm$0.012 &\hspace{0.3cm}
$\pm$0.02 &\hspace{0.3cm} $\pm$0.2 &\hspace{0.3cm} $\pm$0.03
&\hspace{0.3cm}$\pm$0.10 &\hspace{0.3cm}\\    
\hline0.006 &\hspace{0.3cm} 0.25  &\hspace{0.3cm} 4.14 &\hspace{0.3cm}
3.14 &\hspace{0.3cm} 6050  &\hspace{0.3cm} 3.793-3.818 &\hspace{0.3cm}
1.11 &\hspace{0.3cm} 0.69 &\hspace{0.3cm} 0.150 &\hspace{0.3cm} 18.47
&\hspace{0.3cm} 34.2 &\hspace{0.3cm} 1.15 &\hspace{0.3cm} 256.75
&\hspace{0.3cm} 0.074 \\      
 &\hspace{0.3cm} &\hspace{0.3cm} $\pm$0.06 &\hspace{0.3cm}$\pm$0.02
&\hspace{0.3cm} $\pm$50 &\hspace{0.3cm}&\hspace{0.3cm} $\pm$0.06
&\hspace{0.3cm} $\pm$0.03 &\hspace{0.3cm} $\pm$0.040 &\hspace{0.3cm}
$\pm$0.02 &\hspace{0.3cm} $\pm$0.2 &\hspace{0.3cm} $\pm$0.03
&\hspace{0.3cm}$\pm$0.10 &\hspace{0.3cm}\\      
\hline0.004 &\hspace{0.3cm} 0.25 &\hspace{0.3cm} 4.26 &\hspace{0.3cm}
3.15 &\hspace{0.3cm} 6050 &\hspace{0.3cm} 3.793-3.812 &\hspace{0.3cm} 1.11
&\hspace{0.3cm} 0.70 &\hspace{0.3cm} 0.150 &\hspace{0.3cm} 18.49
&\hspace{0.3cm} 34.5 &\hspace{0.3cm} 1.27 &\hspace{0.3cm} 256.79
&\hspace{0.3cm} 0.070\\      
 &\hspace{0.3cm}  &\hspace{0.3cm}$\pm$0.06 &\hspace{0.3cm}
$\pm$0.02 &\hspace{0.3cm} $\pm$50 &\hspace{0.3cm} &\hspace{0.3cm}
$\pm$0.05 &\hspace{0.3cm} $\pm$0.03 &\hspace{0.3cm} $\pm$0.020
&\hspace{0.3cm} $\pm$0.02 &\hspace{0.3cm}$\pm$0.2 &\hspace{0.3cm}
$\pm$0.04 &\hspace{0.3cm}$\pm$0.10 &\hspace{0.3cm} \\ 
\hline 0.002 &\hspace{0.3cm} 0.25  &\hspace{0.3cm} 4.02
&\hspace{0.3cm} 3.13 &\hspace{0.3cm} 6000 &\hspace{0.3cm} 3.793-3.812
&\hspace{0.3cm} 1.06 &\hspace{0.3cm} 0.68 &\hspace{0.3cm} 0.139
&\hspace{0.3cm} 18.46 &\hspace{0.3cm} 34.3 &\hspace{0.3cm} 1.39
&\hspace{0.3cm} 256.81 &\hspace{0.3cm} 0.078 \\    
 &\hspace{0.3cm}   &\hspace{0.3cm} $\pm$0.06
&\hspace{0.3cm} $\pm$0.02 &\hspace{0.3cm} $\pm$50
&\hspace{0.3cm} &\hspace{0.3cm} $\pm$0.10
&\hspace{0.3cm} $\pm$0.05 &\hspace{0.3cm} $\pm$0.014 &\hspace{0.3cm}
$\pm$0.03 &\hspace{0.3cm}$\pm$0.4 &\hspace{0.3cm} $\pm$0.01
&\hspace{0.3cm}$\pm$0.10 &\hspace{0.3cm}\\    
\hline0.004 &\hspace{0.3cm} 0.26 &\hspace{0.3cm} 4.14 &\hspace{0.3cm}
3.14&\hspace{0.3cm} 6050 &\hspace{0.3cm} 3.793-3.816 &\hspace{0.3cm} 1.10
&\hspace{0.3cm} 0.69 &\hspace{0.3cm} 0.156 &\hspace{0.3cm} 18.46
&\hspace{0.3cm} 34.4 &\hspace{0.3cm} 1.27 &\hspace{0.3cm} 256.78
&\hspace{0.3cm} 0.069 \\     
 &\hspace{0.3cm} &\hspace{0.3cm} $\pm$0.06&\hspace{0.3cm}
$\pm$0.02 &\hspace{0.3cm} $\pm$50 &\hspace{0.3cm} &\hspace{0.3cm}
$\pm$0.06 &\hspace{0.3cm} $\pm$0.03 &\hspace{0.3cm} $\pm$0.013
&\hspace{0.3cm} $\pm$0.01 &\hspace{0.3cm} $\pm$0.2 &\hspace{0.3cm}
$\pm$0.07 &\hspace{0.3cm}$\pm$0.10 &\hspace{0.3cm}\\   
\hline 0.004 &\hspace{0.3cm} 0.27  &\hspace{0.3cm}4.14 &\hspace{0.3cm}
3.16 &\hspace{0.3cm} 6100 &\hspace{0.3cm} 3.793-3.805 &\hspace{0.3cm} 1.10
&\hspace{0.3cm} 0.68 &\hspace{0.3cm} 0.171 &\hspace{0.3cm} 18.44
&\hspace{0.3cm} 34.3 &\hspace{0.3cm} 1.20 &\hspace{0.3cm}
256.77&\hspace{0.3cm} 0.068 \\  
 &\hspace{0.3cm}  &\hspace{0.3cm}$\pm$0.06  &\hspace{0.3cm}
$\pm$0.02 &\hspace{0.3cm} $\pm$50 &\hspace{0.3cm} &\hspace{0.3cm}
$\pm$0.02 &\hspace{0.3cm} $\pm$0.01 &\hspace{0.3cm} $\pm$0.015
&\hspace{0.3cm} $\pm$0.02 &\hspace{0.3cm} $\pm$0.2 &\hspace{0.3cm}
$\pm$0.08 &\hspace{0.3cm}$\pm$0.10  &\hspace{0.3cm}\\    
\hline0.004 &\hspace{0.3cm} 0.28 &\hspace{0.3cm} 4.14 &\hspace{0.3cm}
3.15&\hspace{0.3cm} 6100 &\hspace{0.3cm} 3.793-3.795 &\hspace{0.3cm} 1.08
&\hspace{0.3cm} 0.67 &\hspace{0.3cm} 0.172 &\hspace{0.3cm} 18.43
&\hspace{0.3cm} 34.2 &\hspace{0.3cm} 1.20 &\hspace{0.3cm} 256.81
&\hspace{0.3cm} 0.067 \\     
 &\hspace{0.3cm} &\hspace{0.3cm} $\pm$0.06 &\hspace{0.3cm}
$\pm$0.02 &\hspace{0.3cm} $\pm$50 &\hspace{0.3cm} &\hspace{0.3cm}
$\pm$0.02 &\hspace{0.3cm} $\pm$0.01 &\hspace{0.3cm} $\pm$0.015
&\hspace{0.3cm} $\pm$0.02 &\hspace{0.3cm} $\pm$0.2&\hspace{0.3cm} $\pm$0.08
&\hspace{0.3cm}$\pm$0.10 &\hspace{0.3cm}\\   
\hline
\hline
\end{tabular}
\label{tab-bestFit}
\end{center}
} 
\end{table*}

By applying the $\chi^2$ analysis to the models computed at fixed chemical 
composition (Z=0.008, Y=0.25) and stellar mass ($M=4.14M_{\odot}$),  
we find that the effective temperature of the best fit model 
--$T_{e}=6050$$\pm$50 K (see Fig.~1)-- agrees quite well with the 
empirical estimate ($T_{e}=5900$$\pm$250 K). The same outcome applies 
to the other observables, since the agreement between theory and 
observations is better than 1$\sigma$ for the mean radius and the
V-band amplitude, and of the order of 2$\sigma$ for the I-band. 
Note that the uncertainties on the intrinsic 
parameters of the best fit model were estimated according to the step 
in mass value and in effective temperature of the different sets of models.  

To further improve the accuracy of the intrinsic parameters, we computed 
a new set of models at fixed chemical composition and effective temperature
($T_{e}=6050$ K). 
The new models were constructed assuming a step in mass of $\sim$0.06$M_\odot$ 
and once again the luminosity was changed to fit the observed period.  
The $\chi^2$ analysis indicates that the best fit is given by a model 
with a stellar mass of $4.08M_{\odot}$ and a luminosity level
$\log{L/L_{\odot}}$=3.13 dex.  
Data plotted in Fig.~2, where a subset of the computed models is shown, display that theory and observation agree quite
well with each other.

\begin{figure*}
\includegraphics[height=0.8\textheight,width=0.8\textwidth]{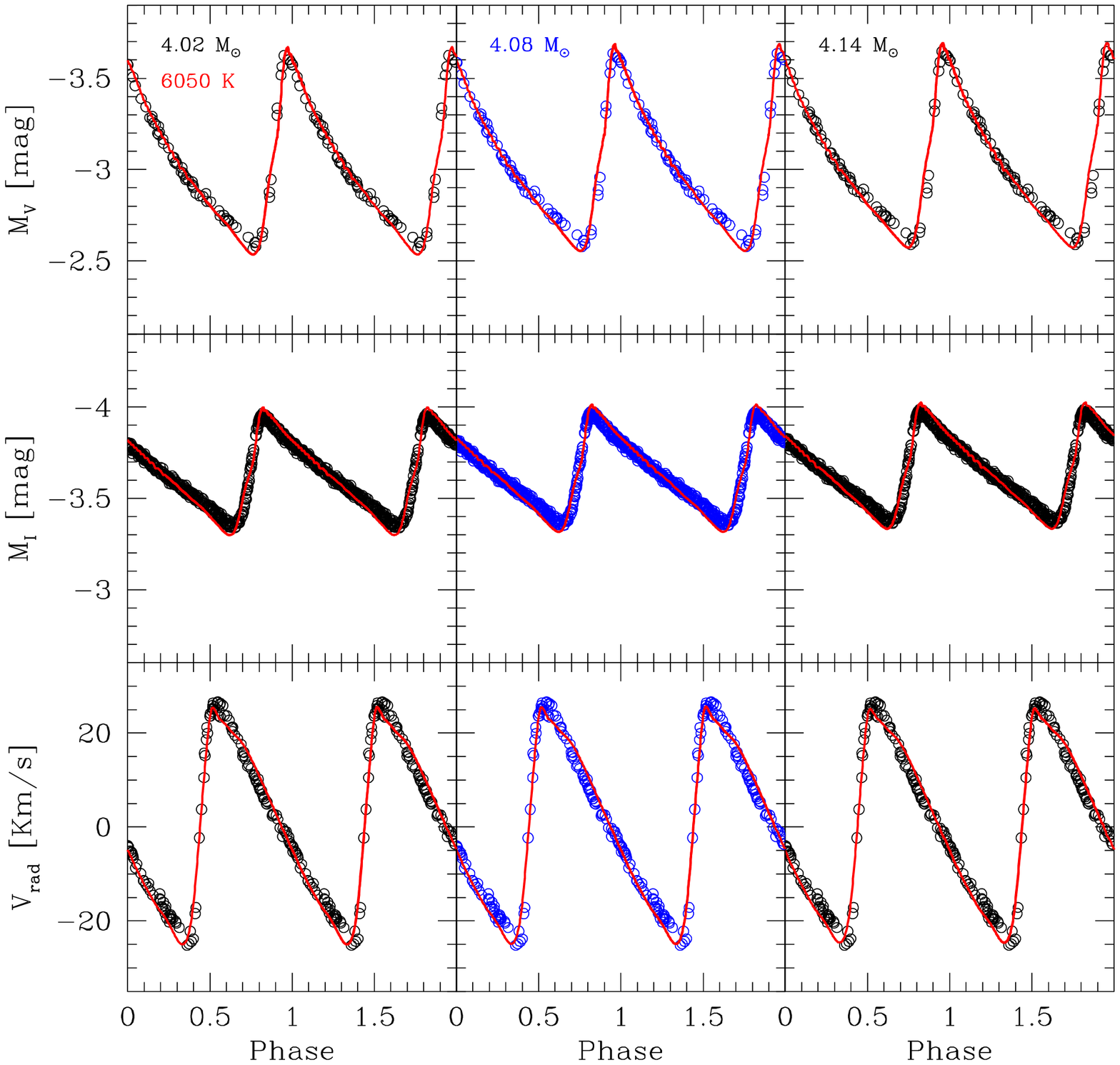}\label{fig2}
\caption{Same as Fig.~1, but the comparison between theory and observations was performed 
at fixed chemical composition (Z=0.008, Y=0.25) and effective temperature ($T_e$=6050 K). 
From left to right the different panels display pulsation models constructed by assuming 
different values of the stellar mass (see labeled values). The blue circles show the 
best fit solution for this set of models.  
}
\end{figure*}

\subsection{Dependence on chemical composition}

In order to constrain the dependence of the best fit solution on
the adopted chemical composition, we computed additional pulsation
models by varying either the metallicity or the helium content.
In particular, we computed new sets of models at fixed helium content 
--Y=0.25--  and three different metal abundances: Z=0.006, Z=0.004 and 
Z=0.002. For each set of models, we performed the same fit to the observed 
curves both at fixed mass and at fixed effective temperature, following 
the same approach we adopted for the models at canonical chemical 
composition (Z=0.008). The best fit solutions for Z ranging from 0.004
to 0.008 are plotted in Fig.~3
and the corresponding stellar parameters for all the metallicities are listed in Table \ref{tab-bestFit}.

\begin{figure*}
\includegraphics[height=0.8\textheight,width=0.8\textwidth]{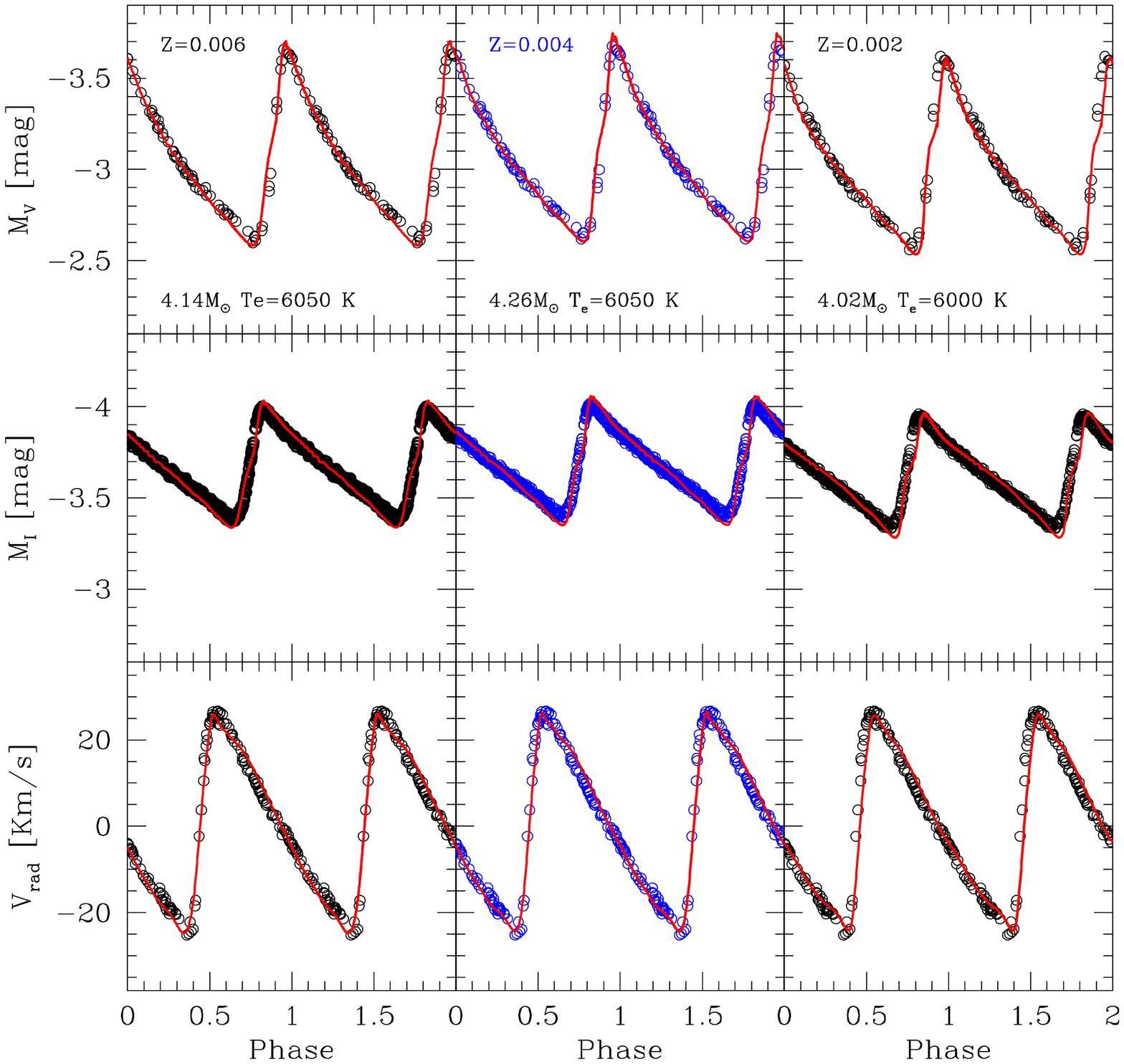} \label{fig3}
\caption{Same as Fig.~1, but for pulsation models constructed by assuming the same helium 
content (Y=0.25) and different values of the metal abundance (see labeled values). 
The stellar mass and the effective temperature of the best fit solution, for each 
set of models, are also labeled. Blue circles display the best fit solution of the 
current set of models.  
}
\end{figure*}

The total rms of the best fit models for the four different metallicities, 
indicate that the global best fit is given by the model with Z=0.004, Y=0.25, 
M=4.26$\pm0.06$ $M_\odot$, $\log{L/L_{\odot}}$=3.15$\pm0.02$ and an effective 
temperature of 6050$\pm50$ K, again in good agreement with the results by 
\citet{pie10} (see also Table~1).\footnote{The match of 
the model at Z=0.002 with the data predicts a barycentric velocity --$\gamma$-- 
at odds with the mean value given by \citet{pie10} and for this reason 
it was not taken into account (see \S 3). However, it was kept in the 
analysis only for the sake of completeness.}. 
Current analysis further confirms the sensitivity of pulsation models to 
the metallicity, and in turn the opportunity to constrain this parameter 
with the comparison between predicted and observed light and radial 
velocity variations \citep{n08}.

To further constrain the impact that chemical composition has on the 
comparison between predicted and empirical observables, we also 
computed new sets of pulsation models at fixed metallicity --Z=0.004-- 
and three enhanced helium contents, namely Y=0.26, 0.27, 0.28.
The dependence of the best fit solution on the helium content is shown 
in Fig.~4. Data plotted in this figure and predicted observables listed 
in Table \ref{tab-bestFit} indicate that the best fit solutions are 
those with He enhanced (Y=0.27, Y=0.28) composition. However, the 
rms values clearly show that pulsation observables are less 
sensitive to changes in He than in metal content. On the basis of 
this evidence and the fact that the set of models with the highest 
He content (Y=0.28) give barycentric velocity --$\gamma$--
slightly larger than the value given by \citet{pie10}, we assume that 
our global best fit model is the one with Z=0.004 and Y=0.27.  

\begin{figure*}
\includegraphics[height=0.8\textheight,width=0.8\textwidth]{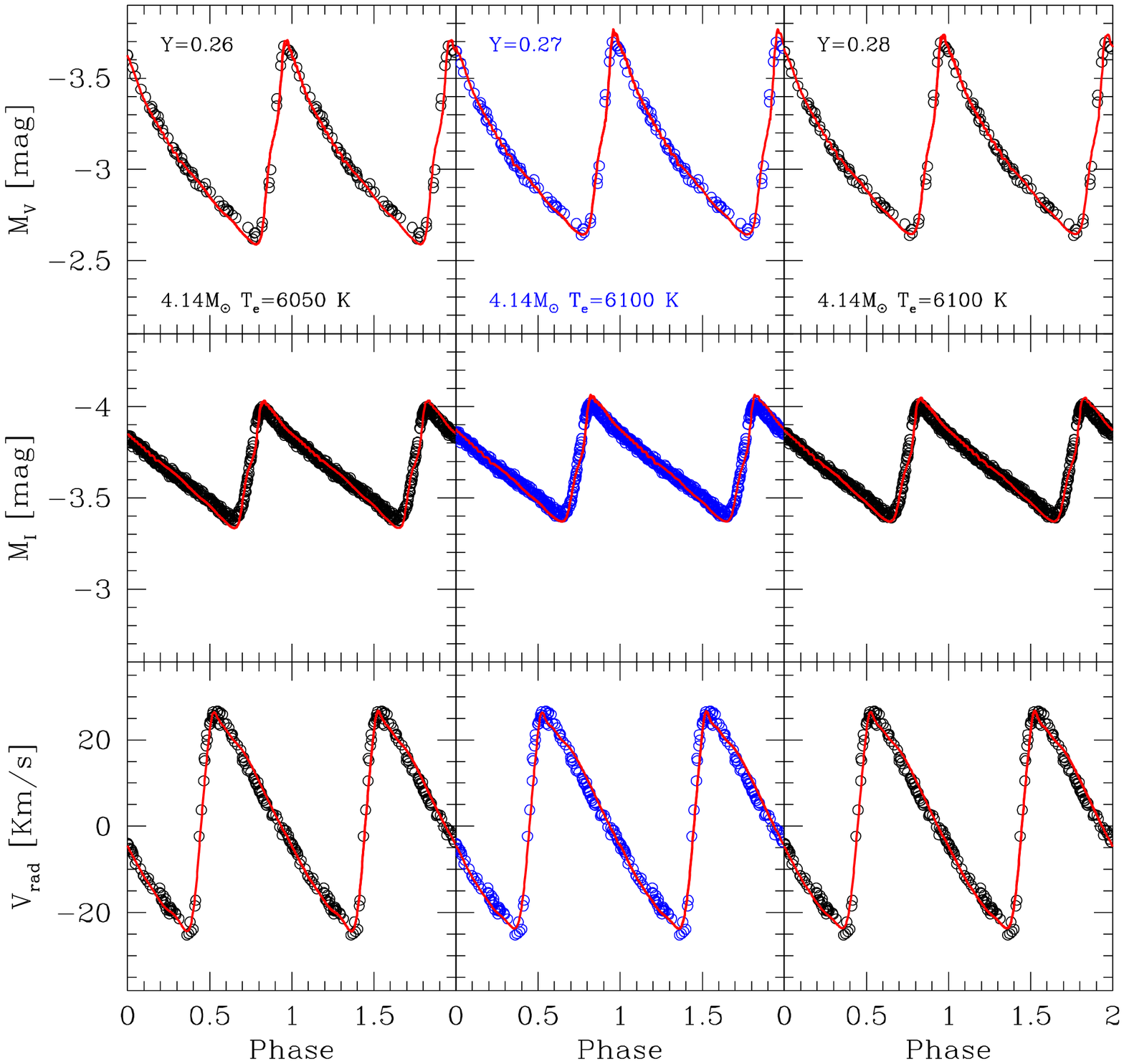} \label{fig4}
\caption{Same as Fig.~3, but for pulsation models constructed by assuming the same metal 
content (Z=0.004) and different helium abundances (see labeled values). The helium  
enhanced (Y$>$0.25) pulsation models cover a broad range in stellar mass 
and in effective temperature. The stellar mass and the effective temperature of the best 
fit solution, for each set of models, are also labeled. Blue circles 
display the global best fit solution.  
}
\end{figure*}

\section{Conclusions and final remarks}

We computed new sets of nonlinear convective pulsation models in
order to match the observed V- and I-band light curve, the radial 
velocity curve and the radius curve of the LMC Cepheid OGLE-LMC-CEP-0227. 
This variable belongs to a detached, double-lined eclipsing binary 
system, and for the first time its dynamical mass was estimated with 
an accuracy better than 1\% \citet{pie10}.
By adopting the canonical chemical composition for LMC Cepheids 
(Z=0.008, Y=0.25), the best fit model gives a value of the Cepheid 
mass --$M=4.08\pm0.06M_{\odot}$-- and of the effective temperature 
--$T_e$=6050$\pm$50 K-- that agree quite well with the observed values. 
This outcome also applies to the mean radius (the difference is smaller 
than 1$\sigma$) and to the pulsation amplitudes (the difference is smaller 
than 2$\sigma$). 
The true distance modulus corresponding to this solution is
$\mu_0$=18.44$\pm$0.02 where the small uncertainty is the only
contribution of the model fitting technique.
By adding a conservative systematic error of 0.1 mag to account for 
several residual uncertainties introduced by the adopted turbulent 
convective model, the static model atmosphere, the input physics, 
together with observational errors, 
we found that current  distance 
is in remarkable agreement with several independent estimates of the
LMC distance in the recent literature
\citep[][]{bono02,mc05,kw06, 
tes07,mol12,m13,p13}.  To provide a firm estimate of the LMC distance
we evaluated the geomtrical 
correction by adopting the model by \citet[][]{vdm02} and we found
that it is 0.062 mag. Therefore, current distance modulus once corrected
for the barycenter of the LMC becomes $\mu_{0,LMC}$=18.50$\pm$0.02 mag,
(d=50.1$\pm$0.5 kpc)  that is in excellent agreement with the very precise
distance recently provided by \citet{p13} using  eight long-period,
late-type double eclipsing binary systems (18.49$\pm$0.05 mag). The same
agreement applies to the distance modulus provided by \citet{pra12} using
evolutionary models (18.53$\pm$0.02 mag).

However, we still lack an accurate measurement of the iron abundance 
of this Cepheid. Therefore, we also investigated possible variations 
of the chemical compositions. We found that the best fit model providing 
pulsation observables that agree quite well with observations and an even 
smaller rms is the model with Z=0.004, Y=0.27, $M=4.14\pm0.06M_{\odot}$, 
$\log{L/L_{\odot}}$= 3.16$\pm$0.02 dex, $T_e$=6100$\pm$50 K and 
$R=34.3\pm 0.2 R_\odot$. This model provides a distance modulus 
corrected  for the barycenter of the LMC  of  $\mu_{0,LMC}$=18.50$\pm$0.02 mag.

Current pulsation mass estimate agrees quite well with the dynamical
mass measurements.  A similar agreement has also been
found concerning the evolutionary masses suggesting either an increase in
the efficiency of convective core overshooting during hydrogen burning
phases \citep{cs11,pra12} and a pulsation driven mass loss
inside the Cepheid instability strip (Neilson et al. 2011).
Current pulsation and evolutionary results are hampered by the fact that 
they are only based on a single object and are affected by 
degeneracy among the different input parameters. The use of 
independent solid observables (period derivative, CNO abundances) 
will provide firm constraints on the actual evolutionary and pulsation 
status of the system (Neilson et al. 2012; Matthews et al. 2012). 

The $p$-factor given by our best fit solution --1.20$\pm$0.08-- is 
consistent with the result obtained for the prototype $\delta$ Cephei 
by \citet{mer05} and with the value suggested by \citet{gro13}
by adopting Galactic Cepheids with individual distances based on 
trigonometric parallaxes. The current estimate is also in very good 
agreement with the new binary best fit solution provided by 
Pietrzynski et al. (2013) and confirms the value predicted by
\citet{nardetto04}  on the basis of Cepheid hydrodynamical models. 
However, the $p$-factor we found is smaller 
than the values recently determined for short-period Cepheids in the LMC 
by \citet{sto11}. The $p$-factor and its possible dependence on the 
pulsation period is still lively debated \citep[][]{gie05,gro13,ngeow12} 
and we need to apply the same analysis to several 
Cepheids before we can rich firm conclusions.  

The reddening estimate is also in very good agreement with recent 
evaluations based on reddening maps (Schlegel et al. 1998) and 
on evolutionary models (Prada Moroni et al. 2012).    
The estimated luminosity level is in remarkable agreement not only 
with the estimate of the binary modelling \citet{pie10}, but also with 
evolutionary prescriptions based on models accounting for mild convective 
core overshooting during central hydrogen burning and canonical 
mass loss rates (see figure 3 in Prada Moroni et al. 2012). This evidence 
indicates a very promising agreement between pulsation and evolutionary 
observables. 
Finally, it is worth mentioning that the nonlinear pulsation approach 
can provide firm constraints not only on the pulsation observables 
--at a few percent level-- but also on the metal abundance. However, 
their sensitivity to the helium content is quite limited.

\acknowledgements
We acknowledge financial support from PRIN INAF 2011 (P.I. M. Marconi)
and from PRIN MIUR 2011 (P.I. F. Matteucci). G.B. thanks ESO 
for support as science visitor. WG and GP thank for support from the 
BASAL Centro de Astrofisica y Tecnologias Afines 
(CATA) PFB-06/2007. Support from Foundation for Polish Science 
(program TEAM) and the Polish National Science Centre 
(program MAESTRO) is also acknowledged.
RS is supported by Polish NCN (grant DEC-2012/05/B/ST9/03932). 
We dedicate this manuscript to the memory of R. Buchler. His 
suggestions and ideas in the field of stellar pulsation will 
be greatly missed.

\pagebreak

\bibliographystyle{apj}

\begin{thebibliography}{}





\bibitem[Bono et al.(1999)]{bms99} Bono, G., Marconi, M., 
\& Stellingwerf, R.~F.\ 1999, \apjs, 122, 167 

\bibitem[Bono et al.(2000)]{b00} Bono, G., Caputo, F., 
Cassisi, S., et al.\ 2000, \apj, 543, 955
 
\bibitem[Bono et al.(2000)]{bms00} Bono, G., Marconi, M., \& Stellingwerf, R.~F.\ 2000, \aap, 360, 245 

\bibitem[Bono et al.(2002)]{bono02} Bono, G., Castellani, V., 
\& Marconi, M.\ 2002, \apjl, 565, L83 

\bibitem[Bono et al.(2002)]{bgmc02} Bono, G., Groenewegen, 
M.~A.~T., Marconi, M., \& Caputo, F.\ 2002, \apjl, 574, L33 

\bibitem[Bono et al.(2008)]{b08} Bono, G., Caputo, F., 
Fiorentino, G., Marconi, M., \& Musella, I.\ 2008, \apj, 684, 102 

\bibitem[Bono et al.(2010)]{b10} Bono, G., Caputo, F., 
Marconi, M., \& Musella, I.\ 2010, \apj, 715, 277 

\bibitem[Brocato et al.(2003)]{bro03} Brocato, E., 
Castellani, V., Di Carlo, E., Raimondo, G., 
\& Walker, A.~R.\ 2003, \aj, 125, 3111 

\bibitem[Caputo et al.(2005)]{c05} Caputo, F., Bono, G., 
Fiorentino, G., Marconi, M., \& Musella, I.\ 2005, \apj, 629, 1021

\bibitem[Cassisi \& Salaris(2011)]{cs11} Cassisi, S., \& Salaris,
  M.\ 2011, \apjl, 728, L43  


\bibitem[Christy(1970)]{Christy70} Christy, R.~F.\ 1970, \jrasc, 
64, 8

\bibitem[Evans et al.(2005)]{Evans2005} Evans, N.~R., Carpenter, 
K.~G., Robinson, R., Kienzle, F., \& Dekas, A.~E.\ 2005, \aj, 130, 789 

\bibitem[Freedman et al.(2001)]{f01} Freedman, W.~L., 
Madore, B.~F., Gibson, B.~K., et al.\ 2001, \apj, 553, 47

\bibitem[Freedman et al.(2011)]{f11} Freedman, W.~L., 
Madore, B.~F., Scowcroft, V., et al.\ 2011, \aj, 142, 192 

\bibitem[Fricke et al.(1971)]{Fricke71} Fricke, K., Stobie, 
R.~S., \& Strittmatter, P.~A.\ 1971, \mnras, 154, 23
 
\bibitem[Gieren et al.(2005)]{gie05} Gieren, W., Storm, J., 
Barnes, T.~G., III, et al.\ 2005, \apj, 627, 224 

\bibitem[Groenewegen(2013)]{gro13} Groenewegen, M.~A.~T.\ 2013, \aap, 550, A70 

\bibitem[Iglesias \& Rogers(1991)]{ir91} Iglesias, C.~A., \& Rogers, F.~J.\ 1991, \apj, 371, 408 


\bibitem[Keller \& Wood(2006)]{kw06} Keller, S.~C., \& Wood, P.~R.\ 2006, \apj, 642, 834 

\bibitem[Luck et al.(1998)]{l98} Luck, R.~E., Moffett, 
T.~J., Barnes, T.~G., III, \& Gieren, W.~P.\ 1998, \aj, 115, 605 


\bibitem[Marconi(2009)]{marconi09} Marconi, M.\ 2009, \memsai, 80, 141

\bibitem[Marconi \& Clementini(2005)]{mc05} Marconi, M., \& Clementini, G.\ 2005, \aj, 129, 2257 

\bibitem[Marconi \& Degl'Innocenti(2007)]{md07} Marconi, M., \& Degl'Innocenti, S.\ 2007, \aap, 474, 557

\bibitem[Marconi et al.(2010)]{m10} Marconi, M., Musella, I., Fiorentino, G., et al.\ 2010, \apj, 713, 615 

\bibitem[Marconi et al.(2013)]{m13} Marconi, M., Molinaro, R.,
  Ripepi, V. et al.\ 2013, MNRAS, 428, 2185

\bibitem[Matthews et al.(2012)]{matt12} Matthews, L.~D., 
Marengo, M., Evans, N.~R., \& Bono, G.\ 2012, \apj, 744, 53 


\bibitem[M{\'e}rand et 
al.(2005)]{mer05} M{\'e}rand, A., Kervella, P., Coud{\'e} du Foresto, V., et al.\ 2005, \aap, 438, L9 

\bibitem[Molinaro et al.(2012)]{mol12} Molinaro, R., et al., 2012,
  ApJ, 748, 69

\bibitem[Moskalik et al.(1992)]{Moskalik92} Moskalik, P., Buchler, J.~R., \& Marom, A.\ 1992, \apj, 385, 685 
\bibitem[Mucciarelli et al.(2011)]{mu11} Mucciarelli, A., Cristallo, S., Brocato, E., et al.\ 2011, \mnras, 413, 837 

\bibitem[Nardetto et 
al.(2004)]{nardetto04} Nardetto, N., Fokin, A., Mourard, D., et al.\ 2004, \aap, 428, 131 


\bibitem[Natale et al.(2008)]{n08} Natale, G., Marconi, M., 
\& Bono, G.\ 2008, \apjl, 674, L93 

\bibitem[Neilson et al.(2011)]{n11} Neilson, H.~R., Cantiello, M., \&
  Langer, N.\ 2011, \aap, 529, L9 

\bibitem[Neilson et al.(2012)]{n12} Neilson, H.~R., Langer, 
N., Engle, S.~G., Guinan, E., \& Izzard, R.\ 2012, \apjl, 760, L18 

\bibitem[Ngeow et
al.(2012)]{ngeow12} Ngeow, C.-C., Neilson, H.~R., Nardetto, N., \& Marengo, M.\ 2012, A\&A, 543, A55

\bibitem[Pietrzy{\'n}ski et al.(2010) ]{pie10} Pietrzy{\'n}ski,
  G., Thompson, I.~B., Gieren, W., et al. 2010, Nature, 468, 542

\bibitem[Pietrzy{\'n}ski et al.(2011)]{pie11} Pietrzy{\'n}ski, G.,
  Thompson, I.~B., Graczyk, D., et al.\ 2011, \apjl, 742, L20 

\bibitem[Pietrzy{\'n}ski et al.(2013)]{p13} 
Pietrzy{\'n}ski, G., Graczyk, D., Gieren, W., et al.\ 2013, \nat, 495, 76 

\bibitem[Prada Moroni et al.(2012)]{pra12} Prada Moroni, 
P.~G., Gennaro, M., Bono, G., et al.\ 2012, \apj, 749, 108 

\bibitem[Romaniello et 
al.(2008)]{r08} Romaniello, M., Primas, F., Mottini, M., et al.\ 2008, \aap, 488, 731 

\bibitem[Saha et al.(2001)]{s01} Saha, A., Sandage, A., 
Tammann, G.~A., et al.\ 2001, \apj, 562, 314 

\bibitem[Schlegel et al.(1998)]{1998ApJ...500..525S} Schlegel, D.~J., 
Finkbeiner, D.~P., \& Davis, M.\ 1998, \apj, 500, 525 
 

\bibitem[Seaton et al.(1994)]{sea94} Seaton, M.~J., Yan, Y., Mihalas, D., \& Pradhan, A.~K.\ 1994, \mnras, 266, 805 

\bibitem[Stobie(1969)]{Stobie69} Stobie, R.~S.\ 1969, \mnras, 144, 511

\bibitem[Storm et al.(2011)]{sto11} Storm, J., Gieren, W.,
  Fouqu{\'e}, P., et al.\ 2011, A\&A, 534, A95

\bibitem[Testa et al.(2007)]{tes07} Testa, V., et al., 2007,
 A\&A, 462, 599

\bibitem[van der Marel et al.(2002)]{vdm02} van der Marel, 
R.~P., Alves, D.~R., Hardy, E., \& Suntzeff, N.~B.\ 2002, \aj, 124, 2639 

\bibitem[Wood et al.(1997)]{was97} Wood, P.~R., Arnold, A., 
\& Sebo, K.~M.\ 1997, \apjl, 485, L25 



\end{thebibliography}

\end{document}